\documentclass[conference]{IEEEtran}
\ifCLASSINFOpdf
\else
\fi
\usepackage[draft]{fixme}
\usepackage[utf8]{inputenc}
\usepackage{algorithm, algorithmicx,algpseudocode}
\usepackage{alltt}
\usepackage{amsfonts}
\usepackage{amsmath}
\usepackage{amssymb}
\usepackage{balance}
\usepackage{caption}
\usepackage{color}
\usepackage{csvsimple}
\usepackage{enumitem}
\usepackage{float}
\usepackage{graphicx}
\usepackage{mathptmx}
\usepackage{url}
\usepackage{hyperref}
\usepackage{multicol}
\usepackage{soul}

\begin{document}

\title{Geodabs – Trajectory Indexing \\ Meets Fingerprinting at Scale}

\author{\IEEEauthorblockN{Bertil Chapuis}
\IEEEauthorblockA{Université de Lausanne\\
bertil.chapuis@unil.ch}
\and
\IEEEauthorblockN{Benoît Garbinato}
\IEEEauthorblockA{Université de Lausanne\\
benoit.garbinato@unil.ch}}

\maketitle

\begin{abstract}

Finding trajectories and discovering motifs that are similar in large datasets is a central problem for a wide range of applications.
Solutions addressing this problem usually rely on spatial indexing and on the computation of a similarity measure in polynomial time.
Although effective in the context of sparse trajectory datasets, this approach is too expensive in the context of dense datasets, where many trajectories potentially match with a given query.
In this paper, we apply fingerprinting, a copy-detection mechanism used in the context of textual data, to trajectories.
To this end, we fingerprint trajectories with \textit{geodabs}, a construction based on geohash aimed at trajectory fingerprinting.
We demonstrate that by relying on the properties of a space filling curve geodabs can be used to build sharded inverted indexes.
We show how normalization affects precision and recall, two key measures in information retrieval.
We then demonstrate that the probabilistic nature of fingerprinting has a marginal effect on the quality of the results.
Finally, we evaluate our method in terms of performances and show that, in contrast with existing methods, it is not affected by the density of the trajectory dataset and that it can be efficiently distributed.

\end{abstract}

\section{Introduction}

The booming trend of ubiquitous computing is massively affecting the volume of data we produce today, in particular via the location traces, or \emph{trajectories}, our smartphones generate.
Such trajectories consist of sequences of locations produced by mobile users via their GPS-capable devices.
In this paper, we address two key problems associated with dense trajectory datasets: \emph{finding similar trajectories} and \emph{discovering common motifs in trajectories}.
These problems are indeed at the heart of many location-based application scenarios, such as car sharing, traffic forecasting, public-transport optimization, etc.
By \emph{dense} trajectory dataset, we mean one containing many (partially) overlapping trajectories.
Consider, for instance, a city like London, congested with roads and streets: the trajectories associated with people traveling through it every day have a high probability of overlapping, at least partially.
This is the type of trajectory data set we consider in this paper.

A common approach to solving these problems consists in splitting the solution into the following two steps:

\begin{enumerate}
\itemsep=0pt
  \item \emph{Select candidate trajectories} by using a \emph{spatial index}
  \item \emph{Compare these trajectories} by using a \emph{distance measure}
\end{enumerate}
 
More precisely, Step~1 consists in querying a spatial index, e.g., a quadtree \cite{finkel1974quad}, an r-tree~\cite{guttman1984r}, a tb-tree~\cite{pfoser2000novel}, a seti-tree~\cite{chakka2003indexing} or a k-d tree~\cite{bentley1975multidimensional},  in order to select candidate trajectories that are similar or that contain common motifs.
Such space-partitioning data structures are typically queried with bounding intervals and sometimes a direction.
Yet they have a major drawback: 
as their bounding strategy are coarse grained, their ability to discriminate long trajectories is not very effective.
This results in many irrelevant trajectories being selected.

Step~2 then consists in using a distance measure, such as the Discrete Fréchet Distance (DFD)~\cite{eiter1994computing} or the Dynamic Time Warping distance (DTW)~\cite{yi1998efficient}, in order to further discriminate the candidate trajectories selected in Step~1.
DFD and DTW give good qualitative results and many systems have adopted them to measure the distance between trajectories.
However, computing~DFD or~DTW for a pair of trajectories of cumulated length~$n$ has a complexity of $O(n^2)$.
Furthermore, discovering similar motifs in a pair of trajectories requires computing DFD for $n^4$~pairs of sub-trajectories~\cite{tang2017efficient}.

\begin{figure*}[t!]
    \centering
    \includegraphics[width=0.99\textwidth]{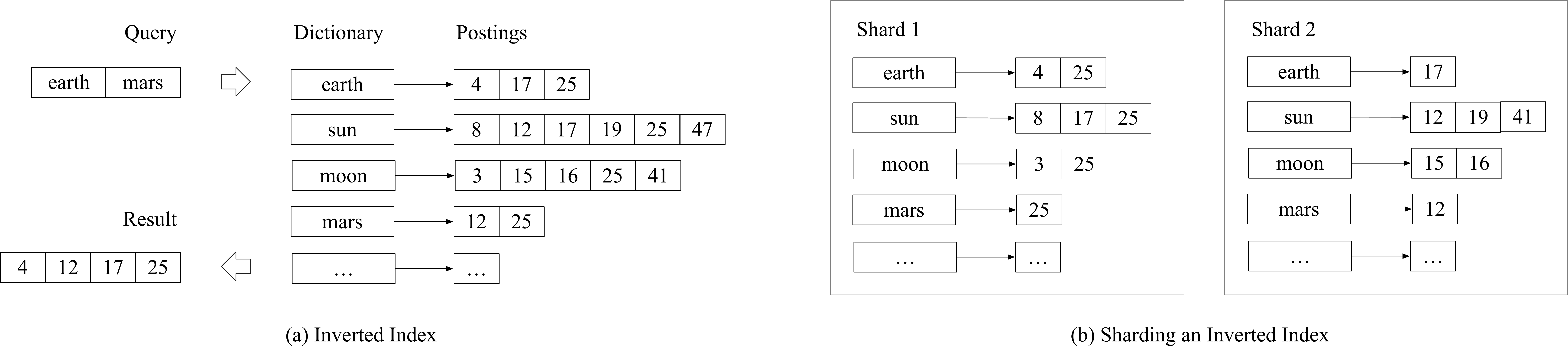}
    \caption{Indexing and Querying Textual Data}
    \label{fig:Indexing}
\end{figure*}

In summary, when faced with a dense set of trajectories, traditional spatial-indexing structures tend to select many irrelevant trajectories, upon which a costly distance measure such as~DFD or~DTW must then be computed.
As a consequence, this combination of techniques  results in serious performance issues when used on dense datasets.

\subsection{Fingerprinting to the Rescue}
\label{sec:motivation}

We argue that the similarity between trajectories and textual data has not been fully exploited.
A text can be seen as a sequence of words, and a trajectory can be seen as a sequence of points.
It is then known that slight variations in word form, e.g., singular vs. plural, conjugation, etc., must be normalized to compare similar but not strictly identical texts.
This is necessary, for instance, to detect plagiarism.
Similarly, minor variations in the location accuracy and sampling rate, which are known to happen when using GPS devices, can compromise the detection of similar trajectories.
Hence, by discretizing trajectories, e.g., by using geohashing~\cite{niemeyer2008geohash}, the negative effect of such minor variations can be mitigated.

A first attempt to exploit this similarity can be found in some geographical information systems that rely on geohashing to create inverted indexes of landmarks.
For example, an open-source search engine called Elastic and adopted by foursquare, relies on this approach.\footnote{\url{https://elastic.co}} 
Google even conceived an alternative to geohash called~S2 that provides the additional guarantee that the surfaces covered by hashes have uniform areas.\footnote{\url{https://s2geometry.io}}
More recently, geohashing has been used for sub-sampling and clustering location traces~\cite{srivatsa2017limits, dewan2017som}.
As of today, however, no research exploits the similarity between trajectories and textual data in terms of their sequentiality, i.e., the fact that sequences of locations are similar to sequence of words.

Extending the analogy between textual data and trajectories is our main contribution in this paper.
More precisely, in the context of textual data, word indexing is known to be ineffective in detecting similarities between large portions of a text, even more so between complete documents.
This is where fingerprinting comes to the rescue, by computing hashes on groups of contiguous words and by using the number of common fingerprints between two documents as a distance measure.
An inverted-index made of fingerprints that point to lists of document identifiers can then be used to efficiently retrieve documents that share some of their content.
By analogy, fingerprinting can be used for finding similar trajectories and discovering similar motifs.
Intuitively, fingerprinting captures the temporal dimension of trajectories by taking into account the ordering of their points.
To our knowledge, the possibilities offered by this observation have not yet been explored.

\subsection{Contribution and Roadmap}

We introduce geodabs, a special kind of fingerprint that can be used for indexing and discovering similarities in dense trajectory datasets.
Geodabs are extracted from trajectories with a fingerprinting algorithm called winnowing~\cite{schleimer2003winnowing}.
Geodabs combine hashing and geohashing to achieve two key properties.
First, hashing addresses the discrimination issue associated with regular spatial indexation techniques.
Second, geohashing enables us to distribute the index accross several nodes in a balanced fashion.
As a result, geodabs can be used to create effective and scalable trajectory indexes.

The remainder of the paper is organized as follows.
We formally introduce the problems addressed in this work, together with some basic definitions, in Section~\ref{problem}.
Then, in Section~\ref{background}, we provide the background required to understand our approach and we discusse related work.
In Section~\ref{sec:solution}, we describe geodabs, our fingerprinting based-solution.
In Section~\ref{sec:trajectory-normalization}, we shows how normalization affects two key measures in information retrieval, namely \emph{precision} and \emph{recall}.
Finally, in Section~\ref{sec:evaluation}, we evaluates geodabs both in terms of efficiency and effectiveness.

\section{Trajectory-based querying}
\label{problem}

In this section, we introduce some basics definitions and formally state the problem addressed in this paper.

\subsection{Moving Objects, Trajectories and Distances}

Every object located on earth has a real position that can be expressed with a latitude-longitude point $p = (\varphi, \lambda)$.
The real position~$p$ of a moving object at time~$t$ can be expressed with a continuous-time function $P(t) = p$.
In practice, the points forming a trajectory are obtained with some GPS-tracking device, which reduces the continuous-time function $P$ to a discrete trajectory $S$ with a certain degree of accuracy.
Hence, formally, a trajectory is modeled as a sequence of points $S = \langle s_1, ..., s_n \rangle$. 
The length of a trajectory is denoted $\mathit{length}(S)$.
A motif (sub-trajectory) of $S$ is denoted $\bar{S}$.

Several methods enable us to compute the distance between two trajectories.
Depending on the method, the distance has different scales and meanings.
Here, we generalise this idea with the distance function $\delta(S_i, S_j) = d$, with $d \in \mathbb{R}^+_0$.
The smaller the distance between a pair of trajectories, the greater their similarity is.

\subsection{Finding Similar Trajectories and Motifs}

Here, we address the problems of \emph{finding similar trajectories} and of \emph{discovering common motifs in trajectories}.
As we target dense trajectory datasets, we express these problems in terms of ranked retrieval, i.e., many trajectories are expected to match a given query.
In addition, ranked retrieval also implies sorting the matching trajectories according to some criterion, in order to place those most relevant early in the result list. 
With this in mind, we can now formally define the two problems addressed in this paper.

\subsubsection{Finding similar trajectories}
Given 
a trajectory dataset $D = \{S_1, ..., S_n\}$, 
a trajectory $S_q \notin D$, 
a distance function~$\delta$ and
a distance~$\Delta_{max}$,  
the problem consists in returning an ordered set $R \subseteq D$ where $S \in R \Rightarrow \delta(S_q,S) \leq \Delta_{max}$.
The ordering in~$R$ is then defined as follows: 
$\forall S_i \in R$ and $\forall S_j \in R$ with $i < j$, 
we have that
$\delta(S_q,S_i) \leq \delta(S_q,S_j)$.
That is, trajectories in~$R$ are at a maximum distance~$\Delta_{max}$ from trajectory query~$S_q$ and they are ordered by their distance with respect to~$S_q$.

\subsubsection{Discovering common motifs in trajectories}
Given two trajectories $S_i$ and $S_j$, 
a distance function~$\delta$ and
a length~$l$,
the problem consists in returning a pair of motifs $(\bar{S}_i,\bar{S}_j)$ such that 
$\mathit{length}(\bar{S}_i) = \mathit{length}(\bar{S}_j) = l~\wedge \not\exists (\bar{S'}_i,\bar{S'}_j)$
for which 
$\delta(\bar{S'}_i,\bar{S'}_j)~<~\delta(\bar{S}_i,\bar{S}_j)$.
That is, among all the pairs of motifs of~$S_i$ and~$S_j$ of length~$l$, $(\bar{S}_i,\bar{S}_j)$ is the one with the smallest distance between them.

\section{Background and related work}
\label{background}

\begin{figure*}[t]
    \centering
    \includegraphics[width=0.99\textwidth]{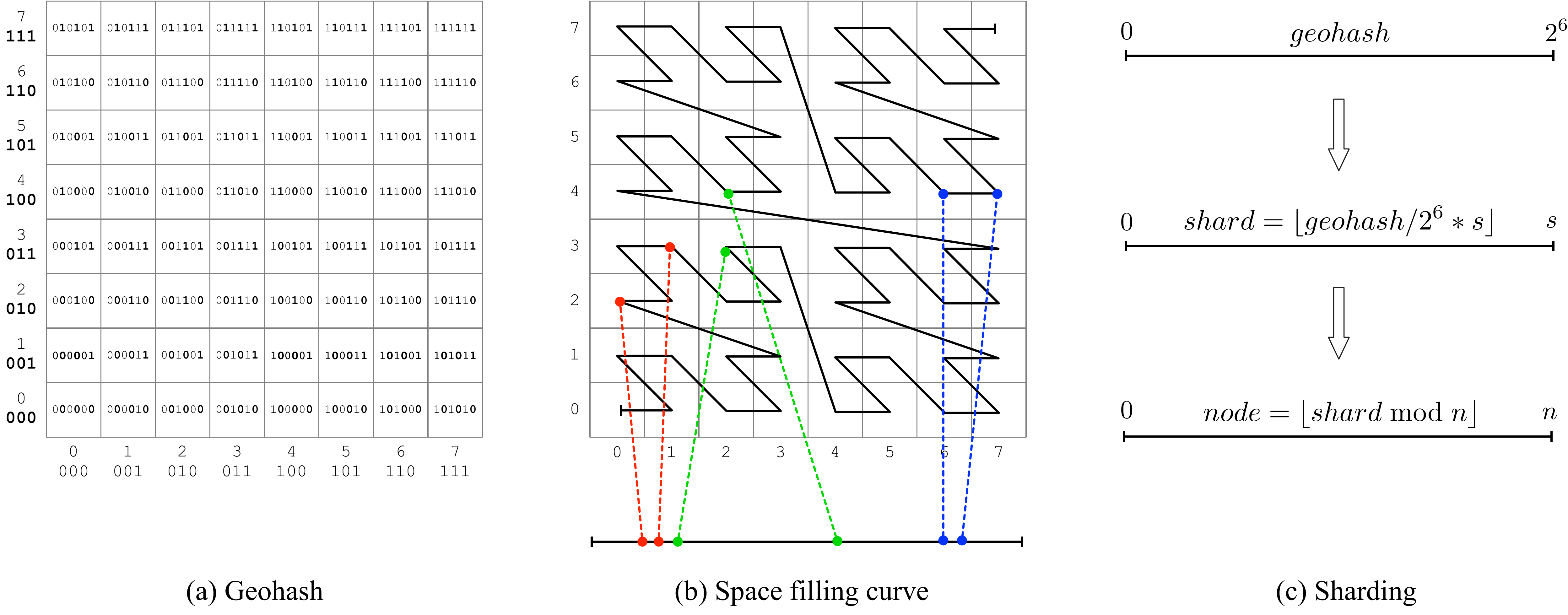}
    \caption{Geohash, space-filling curve and sharding}
    \label{fig:Geohash}
\end{figure*}

\subsection{Information Retrieval}

\subsubsection{Boolean Retrieval}
\label{sec:boolean-retrieval}
As highlighted in Figure \ref{fig:Indexing} (a), in its simplest form, an inverted index is usually composed of terms that point to collections of document identifiers called postings lists~\cite{Manning:2008:IIR:1394399}. 
Boolean queries can then be used to retrieve all the documestnts that contain a set of words.
Optionally, a posting list can also contain the position of the term in the document.
This positional information can then be used to search for sub-sequences in documents.
However, when searching for long sub-sequences of terms, this approach showcases poor performances.
In the context of trajectory indexing, we replace the terms of the inverted index with features, called geodabs, extracted from trajectories.

\subsubsection{Ranked Retrieval}
In ranked retrieval, many records match with the query specified by the user, and it is common to rank the results according to a similarity measure.
Therefore, the user can begin by considering the most relevant results and then decide to retrieve the remaining ones if necessary.
In the context of textutal data, the union and the intersection of two sets of words $F$ and $G$ can be used to derive relevant similarity measures.
For example, the Jaccard coefficient $J(F,G)$ is commonly used to gauge the similarity between texts and rank results~\cite{Manning:2008:IIR:1394399}.
Interestingly, the Jaccard distance $d_J(F,G)$ expressed in Equation \ref{eq:jaccard} is complementary to the Jaccard coefficient and is proven to obey the triangular inequality \cite{kosub2016note}.
Therefore, this distance can be used in conjunction with an index of pre-computed distances to efficiently prune candidates.
In our context, we use the Jaccard distance to implement function $\delta(S_q,S_j)$ and rank the trajectories retrieved from the inverted index.

\begin{equation}
\label{eq:jaccard}
d_J(F,G) = 1 - J(F,G) = 1- {{|F \cap G|} \over {|F \cup G|}}
\end{equation}

\subsubsection{Normalization}
It is worth noting that, in many cases, terms can be different but convey similar semantics or meaning. 
In general, the process of mitigating these differences is referred to as normalization~\cite{Manning:2008:IIR:1394399}.
For example, a common normalization technique consists in using equivalence classes for synonyms.  
In the context of trajectory indexing, we refer to normalization as a function $N(S) = S'$, where $S$ and $S'$ are  sequences of points.

\subsubsection{Sharding}

When an index becomes very large, it might not fit on a single computer anymore.
As illustrated in Figure~\ref{fig:Indexing}~(b), sharding the index across the nodes of a cluster becomes necessary.
Here, the idea is to route documents to specific shards in order to spread the load throughout the cluster.
At query time, all the shards might need to process a query to compute the result.
Therefore, given the terms specified in a query, a good sharding strategy tries to minimize the number of shards that need to be contacted.
Our solution effectively addresses this issue.

\subsection{Fingerprinting}

As mentioned in Section \ref{sec:boolean-retrieval}, searching for long sub-sequences in textual data by using words and positional information is not very efficient.
In practice, an inverted-index aimed at searching for sub-sequences is usually populated with a different class of terms, referred to as fingerprints~\cite{broder1997resemblance, manber1994finding, heintze1996scalable, brin1995copy}.
Fingerprints usually correspond to a sub-set of the hash sums obtained by hashing the n-grams of a document~\cite{schleimer2003winnowing}.
A n-gram is a sequence of $n$ contiguous items, i.e., $n$ words in the context of textual data.
A fingerprint usually corresponds to the hash sum $h \in \mathbb{R}$ obtained by hashing a n-gram.
As the number of n-grams for a given text can be very large, a common practice consists in retaining the subset of fingerprints that satisfy the condition $h \bmod p = 0$, where $p$ is a fixed sampling constant.
The extracted fingerprint can then be used as terms in the inverted index.
Furthermore, given two sets of fingerprints, their similarity can easily be derived by the Jaccard coefficient.
In our context, we refer to fingerprinting as a function $W(S) = F$, where $S$ is a trajectory and $F$ is an ordered set of fingerprints.

\subsection{Geohashing}

A function that produces geohashes maps a point $p$ to a sequence of bits $b$ that repeatedly bisect space up to a desired depth $d$ that defines the precision of the geohash~\cite{niemeyer2008geohash}.
In the case of a latitude/longitude space, the first subdivision usually occurs on the longitude axis, the second on the latitude axis and the process is repeated up to depth $d$.
Figure~\ref{fig:Geohash}~(a) illustrates this subdivision for a depth $d = 6$, where two interleaved sequences of three bits are respectively dedicated to the subdivision of the longitude and latitude axes.
Every geohash covers a delimited area on earth and, given a set of points $\{p_1, p_2, ..., p_n\}$, it is relatively easy to find the highest precision geohash that overlaps with the whole set.
Hereafter, we formally refer to such an overlapping geohash with the function $geohash(\{p_1, p_2, ..., p_n\}) = b$.

As highlighted in Figure \ref{fig:Geohash} (b), the ordered list of hashes obtained by subdividing the space with a geohash function can be represented as a z-order space-filling curve.
Interestingly, when two points are close to each other on a space-filling curve, then they are close to each other in the latitude/longitude space. 
However, the reverse is not necessarily true since: two points near each other in the latitude/longitude space might be far from each other on the space-filling curve.

Then, as illustrated in Figure \ref{fig:Geohash} (c), the space-filling curve can be used to shard a spatial index across the nodes of a cluster.
Two steps characterize the sharding strategy pictured here.
First, geohashes are mapped to shards in a locality preserving way, i.e., geohashes near each other on the space-filling curve are placed on the same shard.
Second, the shards are mapped to nodes with a modulo operation that breaks locality to improve the balance of the index across the nodes.

\section{Fingerprinting with Geodabs}
\label{sec:solution}

We now introduce geodabs, a construction that combines geohashing and hashing to fingerprint trajectories.
Our motivation for introducing geodabs is based on two key reasons.
First, as illustrated in Figures \ref{fig:Geohash} (b), the correspondence between the distance on the space-filling curve and the latitude/longitude space can be used as a means to shard and distribute the index. 
Sharding with geohash on the space-filling curve guarantees that the locality of the index will be preserved and that a minimal number of shards will be contacted to answer a query.
Second, the fingerprints of a trajectory should capture the notion of order present in trajectories.
Given a sequence of points, geohashes only capture the area that overlap with these points.
Therefore, we use regular hashing to address this issue.

\begin{figure}[h!]
    \centering
    \includegraphics[width=0.45\textwidth]{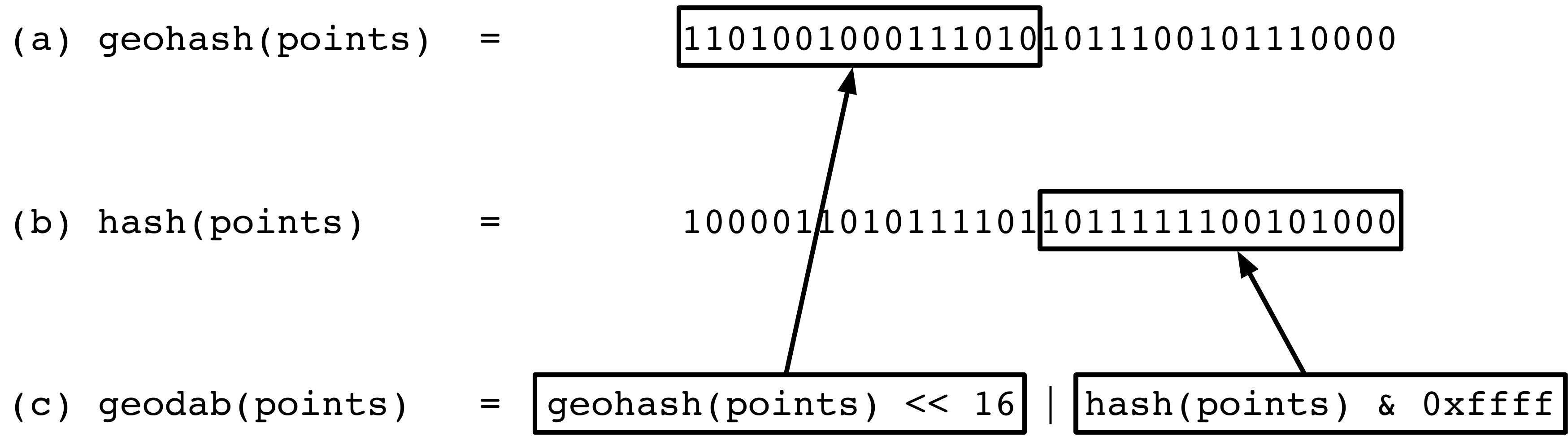}
    \caption{Construction of a geodab}
    \label{fig:Geodabs}
\end{figure}

Figure~\ref{fig:Geodabs} illustrates how geodabs are constructed. 
(a) Given a sequence of points, we first compute the geohash that overlaps with the whole sequence.
The geohash acts as a prefix that naturally distributes the geodabs on the space-filling curve according to the location of the points.
(b) We compute a hash that is sensitive to the ordering of the points.
This hash acts as a suffix for the geodab that discriminates among sequences of points, according to their path and their ordering.
(c) We merge the two hashes, here encoded on $32$ bits, using bitwise operators.
The length of the prefix can easily be adjusted, depending on the desired space partitioning.

\subsection{Trajectory Fingerprinting and Indexing}

When comparing sequences, finding similarities by sampling fingerprints is probabilistic and does not guarantee the detection of all the similarities.
Winnowing is a special fingerprinting algorithm that address this problem and comes with additional guarantees~\cite{schleimer2003winnowing}.
First, it ensures that matches shorter than a user defined lower-bound $k$ are considered as noise.
To satisfy this guarantee, winnowing only considers hashes of k-grams.
Second, it ensures that at least one k-gram is detected in any common sequence of size greater or equal to an upper-bound $t \geq k$.
To satisfy this guarantee, the algorithm defines a window of size $w = t - k + 1$ that slides over the sequence of k-gram hashes.
For each window, the algorithm selects the minimum hash value or the right-most minimum hash value if the same hash appears more than once in the window.
As the dataset densifies, the upper threshold can be used to reduce the number of fingerprints extracted from queries in order to set the efficiency/effectiveness tradeoff.

\begin{figure}[t!]
    \centering
    \includegraphics[width=0.4\textwidth]{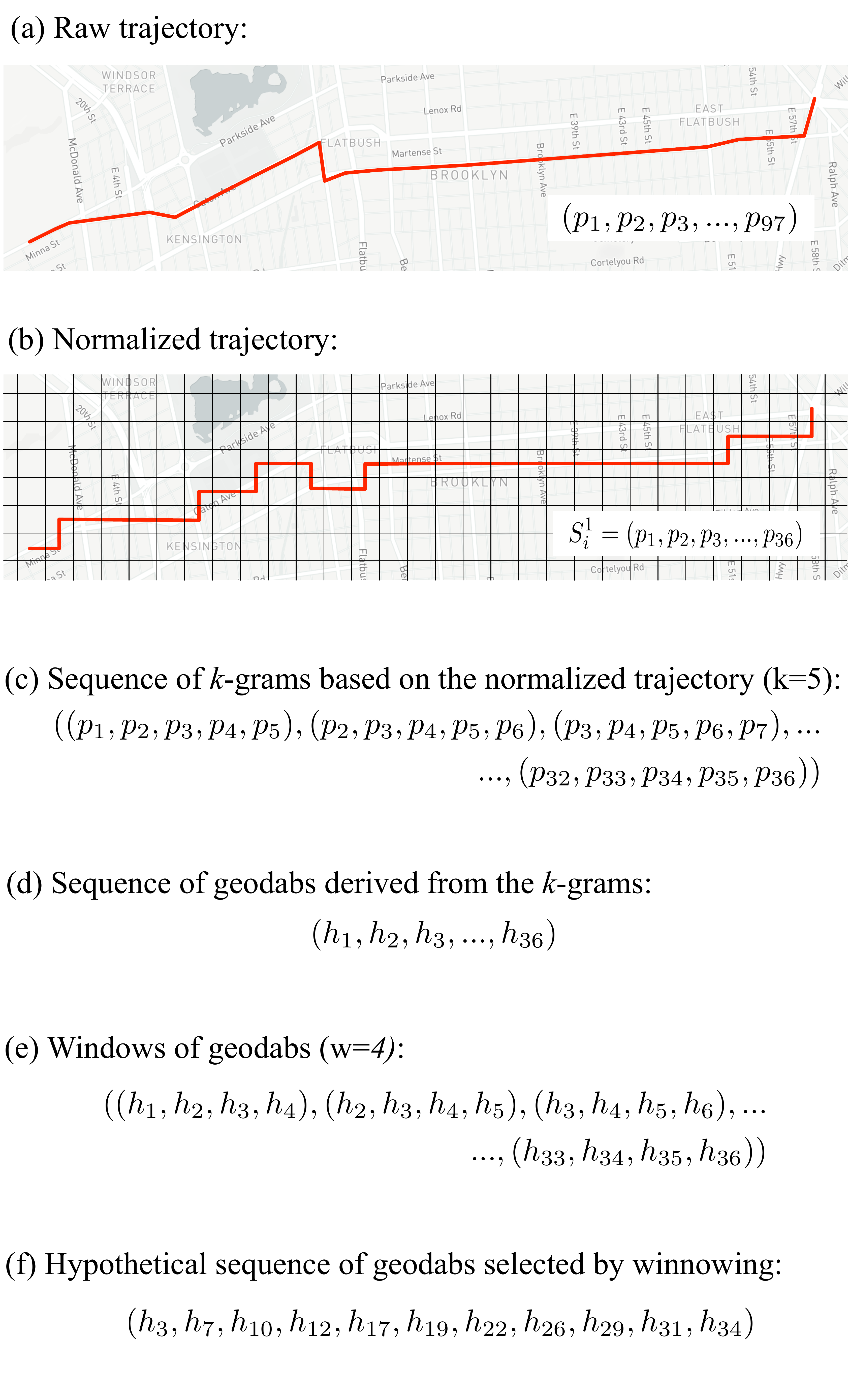}
    \caption{\vspace{-12pt}Trajectory Winnowing}
    \label{fig:Winnowing}
\end{figure}

Figure~\ref{fig:Winnowing} shows how the steps described in~\cite{schleimer2003winnowing} can be adapted to extract fingerprints from trajectories.
(a) The raw trajectories can showcase different sampling rates, (b) it is therefore necessary to normalize them.
In Section \ref{sec:trajectory-normalization}, we discuss how normalization can be used to make trajectories converge to similar sequences.
(c) The sequence of k-grams can then be computed and (d) the corresponding hashes can be derived.
(e) The sliding window of size $w$ can then be used to (f) select the hashes that constitute the fingerprints of the trajectory.
The resulting fingerprints can then be used as terms in an inverted index, where posting lists are filled with trajectory identifiers.

Algorithm~\ref{alg:winnowing} shows in more detail how winnowing can be implemented for fingerprinting trajectories.
An optimised version of this algorithm relies on circular buffers and rolling hash functions for iterating over k-grams of points and windows of hashes.
In contrast with normalized documents that often contain thousands of words, normalized trajectories are relatively short sequences of points.
As we did not notice a significant performance gain, we dropped this optimization.

\begin{algorithm}[t!]
\caption{Geodabs extraction by winnowing}
\label{alg:winnowing}
\begin{algorithmic}[1]
    \State \textbf{Input:} trajectory $S$, lower-bound $t$, upper-bound $k$
    \State \textbf{Output:} Geodabs $G$
    \State $C \gets (\varnothing)$ \Comment{Sequence of candidate hashes}
    \For{$i \gets 1$ to $|S| - k$} \Comment{Iterate over k-grams}
    	\State $g \gets geodab(S_{i, i+k})$ \Comment{Compute k-gram geodabs}
    	\State $C \gets C \parallel (g)$ \Comment{Add geodabs to candidates}
  	\EndFor
  	\State $G \gets \{\varnothing\}$ \Comment{Set of winnowed geodabs}
  	\State $w \gets t - k + 1$ \Comment{Window size}
  	\For{$i \gets 1$ to $|C| - w$} \Comment{Iterate over windows}
  		\State $m \gets i$ \Comment{Initialize minimum geodabs index}
  		\For{$j \gets i+1$ to $i+w$} \Comment{Iterate over the window}
  			\If{$C_j <= C_m$} \Comment{Select right most minimum}
  				\State $m = j$ \Comment{Set new minimum}
  			\EndIf
  		\EndFor
    	\State $G \gets G \cup \{C_m\}$ \Comment{Add minimum to geodabs}
  	\EndFor
  	\State \textbf{return} $G$
\end{algorithmic}
\end{algorithm}

In our implementation, we use roaring bitmaps to represent the sets of fingerprints $F$~\cite{lemire2017roaring}.
Any set made of integers can be represented with bitmaps.
Given two bitmaps, their intersection, union, or difference can be computed very efficiently with bitwise operations.
Roaring bitmaps are fast memory-efficient bitmaps that outperform most existing techniques.
The fingerprints generated by the trajectory-winnowing algorithm are used as terms in the inverted index.
Each entry of the postings lists contain a reference to the raw trajectory and a reference to the trajectory bitmap.
As a result, when querying the index, the bitmap of the query can be compared to the bitmap of the result in order to gauge their similarity.

\section{Trajectory Normalization}
\label{sec:trajectory-normalization}

In this section, we show that a sound normalization procedure can be applied to trajectories.
In the context of textual data, normalization relies on semantic rules and linguistic techniques to find equivalence classes for terms.
For example, such techniques include case-folding, true-casting, stemming, lemmatization, and the identification of stop-words.
These techniques remove superficial differences in textual data.
Normalization in the context of trajectories differs, but the notion of equivalence classes remains the same, as highly similar trajectories should converge toward similar sequences of points.
In this section, we introduce two normalization methods and show the extent to which trajectory data should be normalized.

\subsection{Normalizing with Geohash}

A simple normalization technique consists in mapping the points of a trajectory to a sequence of geohashes at a constant depth $d$.
The resulting geohashes can then be cleaned from consecutive duplicates and converted back to sequences of points.
This approach is very lightweight and removes most irrelevant differences in trajectories.
If two trajectories follow the edges of one or several geohashes, the resulting sequences of normalized points could be different.
However, as it will be demonstrated in Section~\ref{sec:evaluation}, it does not really affect the quality of the results.

\subsection{Normalizing with Map Matching}

Another normalization technique consists in mapping trajectories to an existing road network~\cite{newson2009hidden}.
This approach, called map matching, can give very good results, especially if the moving entity at the origin of a trajectory is known to be constrained to a road network.
Most recent map-matching algorithms rely on the Viterbi algorithm to compute a match~\cite{goh2012online}.
For each point of a trajectory, the idea is to first retrieve a set of matching nodes on a road network within a certain radius. 
The Viterbi algorithm then computes the most probable sequence of nodes on the road network~\cite{newson2009hidden}.
This approach is computationally costly, but this price is paid only at the creation of the index. 
When searching for similarities, the query trajectory has to be normalized in a similar way.

\subsection{Extent of the Normalization}

\begin{figure}[t!]
    \centering
    \includegraphics[width=0.45\textwidth]{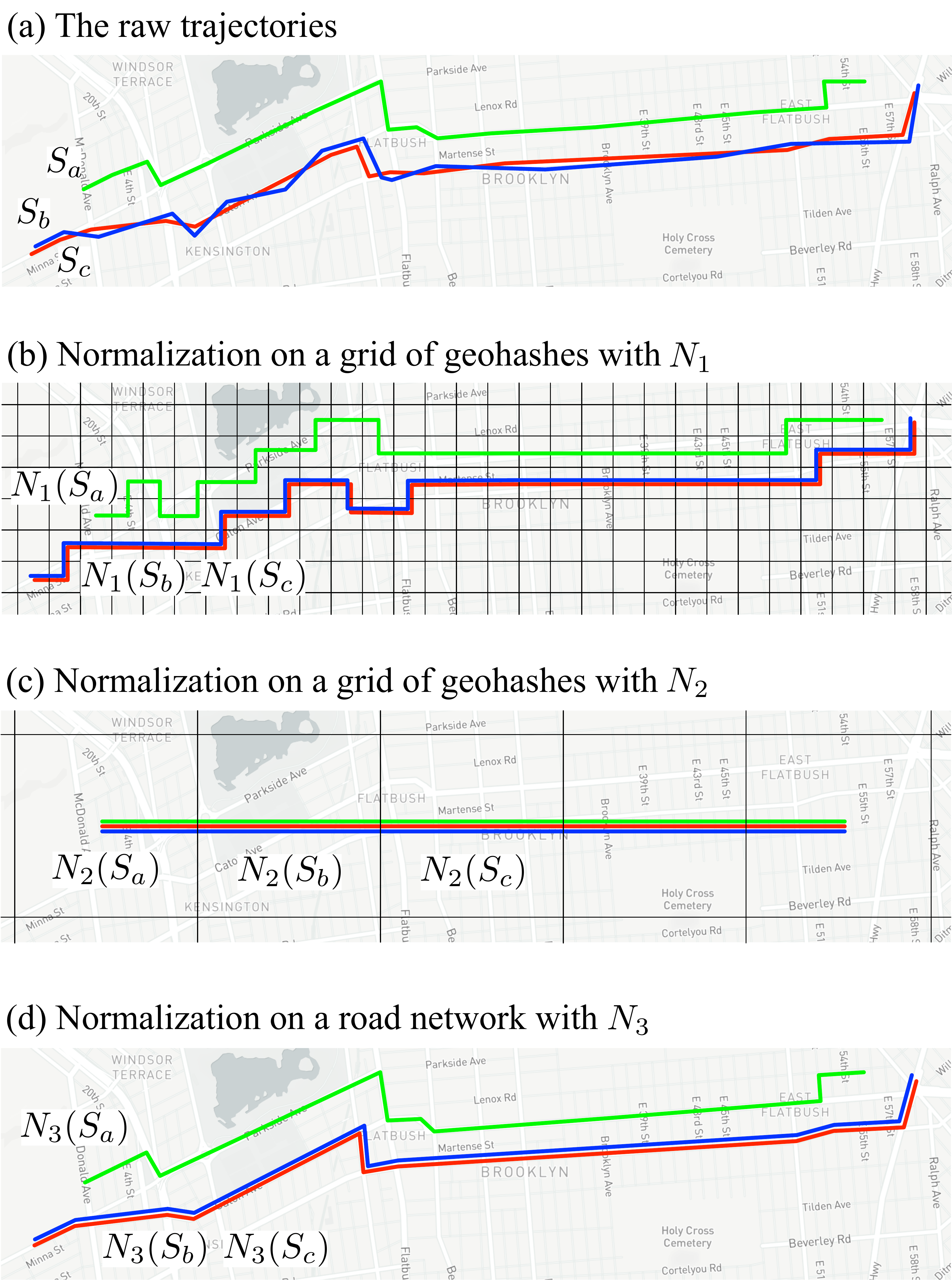}
    \caption{Trajectory Normalization}
    \label{fig:Normalization}
\end{figure}

In the context of textual data, normalization usually relies on simple intuitions.
Unfortunately, when dealing with trajectories, it is difficult to rely on the same intuitions. 
In this section, we show how the measures introduced in~\cite{chapuis2017dss} can be applied to trajectory normalization.

As stated previously, with normalization, trajectories should converge toward similar sequences of points.
Therefore, a good normalization function $n$ applied to a pair of similar trajectories $S_a$ and $S_b$ should satisfy the property $J(W(N(S_a)),W(N(S_b)))>J(W(S_a),W(S_b))$.
Focusing solely on this property, however, leads to aggressive normalization functions that make every trajectories converge to the same output.

In order to address this issue, we start by recalling two key effectiveness measures in information retrieval.
First, precision measures the fraction of selected items that are relevant.  
More formally, given the number of relevant items retrieved $tp$ (true positive) and the number of irrelevant items retrieved $fp$ (false positive), $precision = tp / (tp + fp)$.
Second, recall measures to the fraction of relevant items that are selected.
In other words, given the number of relevant items retrieved $tp$ and the number of relevant items that have not been retrieved $fn$ (false negative), $recall = tp / (tp + fn)$.

In Figure \ref{fig:Normalization}, we illustrate the effect of two hypothetical normalization functions $N_1$ and $N_2$ on precision and recall.
We first introduce a dataset that contains the raw trajectories $S_a$, $S_b$.
We also assume a query trajectory $S_q$, which has one relevant result $S_b$ in the dataset.
Figure \ref{fig:Normalization} (a) depicts an inverted index built with the raw trajectories and queried with $S_q$.
Without normalization, this index would return no relevant result
Both precision and recall would therefore be equal to $0$.
Figure \ref{fig:Normalization} (b) depicts an inverted index built with the normalization function $N_1$ and queried with $S_q$.
This index would return one relevant results.
Thus, precision would be $1 / (1 + 0)$ and recall $1 / (1 + 0)$.
Figure \ref{fig:Normalization} (c) depicts an inverted index built with a more aggressive normalization function $N_2$.
In this case, $S_q$ would return two results, precision would drop to $1 / (1 + 1)$ and recall would remain stable at $1 / (1 + 0)$.
As long as all the true positives are all included in the result set, recall remains high.
However, in the context of geodabs, an aggressive normalization function could over simplify trajectories.
In this case, recall would start dropping, especially if the normalized sequence of points is shorter than the noise threshold specified by the winnowing algorithm.
Observing the evolution of precision and recall is therefore a good way to determine the extent of the normalization.
In ranked retrieval, this is precisely the aim of a precision and recall (PR) curve~\cite{Manning:2008:IIR:1394399}.
In section \ref{sec:configuration}, we empirically show how a PR curve can be used to find the best parameters for a normalization function.

In conclusion, as long as precision and recall improves, the extent to which a trajectory is normalized can be increased.
In contrast, a drop in precision or recall clearly indicates that the fingerprints do not capture what characterizes and differentiates the sequences of points anymore.
Hence, to identify the optimal extent of a normalization function, the evolution of precision and recall can be observed on a sample dataset.

\section{Evaluation}
\label{sec:evaluation}

\begin{figure}[t!]
    \centering
    \includegraphics[width=0.4\textwidth]{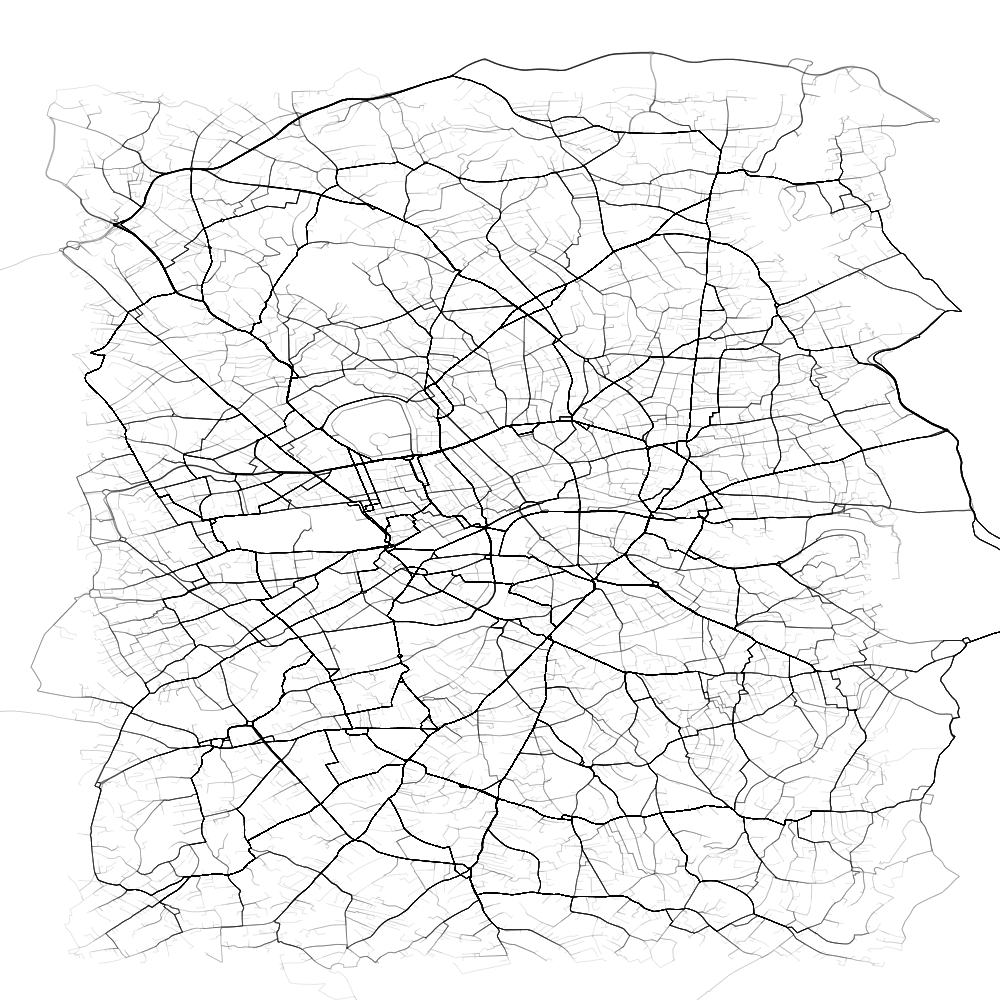}
    \caption{Routes used to generate the trajectory dataset}
    \label{fig:dataset}
\end{figure}

\begin{figure}[t!]
    \centering
    \includegraphics[width=0.35\textwidth]{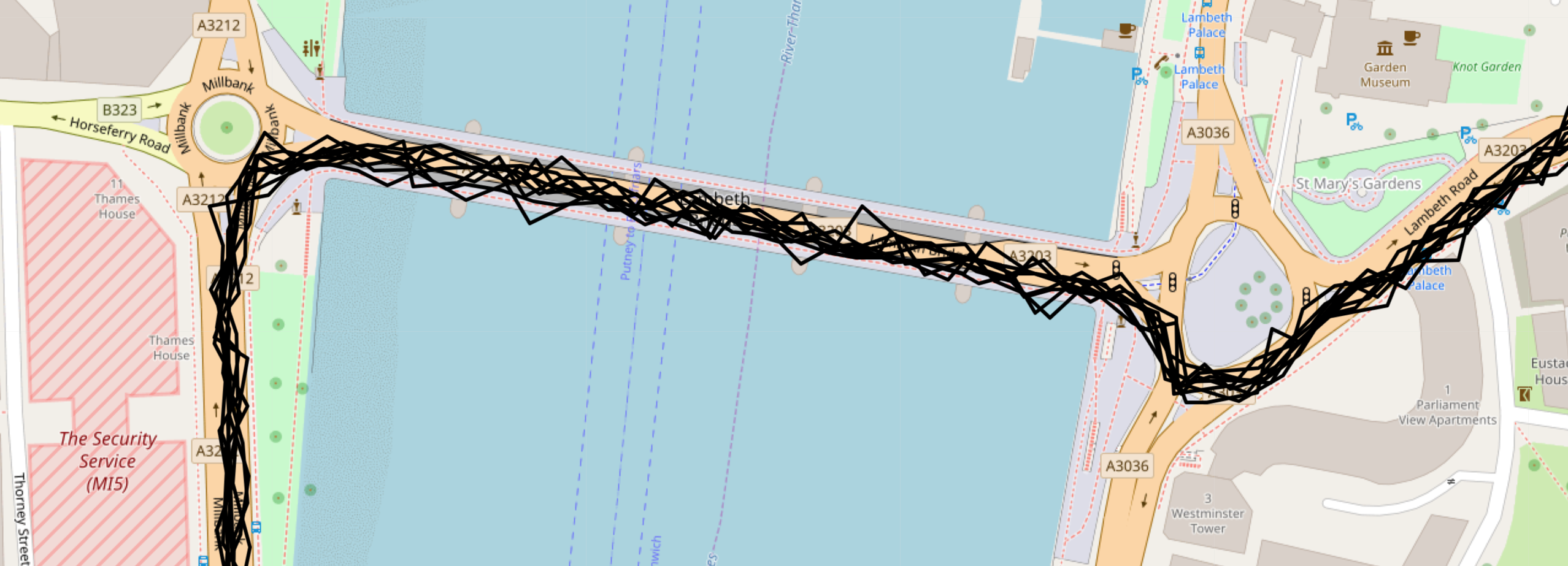}
    \caption{Similar trajectories extracted form the dataset}
    \label{fig:dataset-excerpt}
\end{figure}

\begin{figure*}[t!]
	\captionsetup{justification=centering}
	\begin{minipage}[t]{0.33\linewidth}
		\centering
    	\includegraphics[width=1\textwidth]{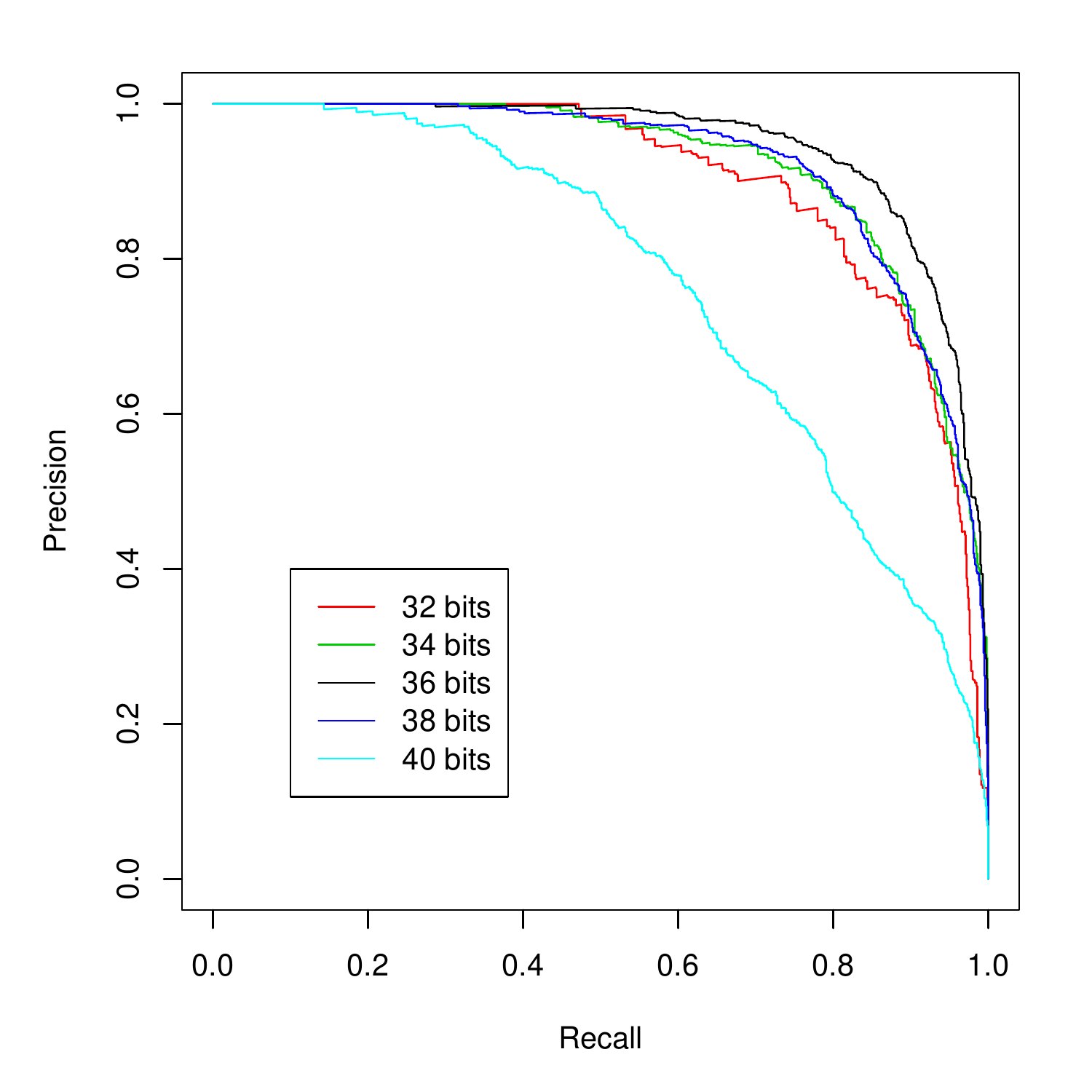}
    	\caption{Verifying configuration parameters with a PR curve}
    	\label{fig:config}
	\end{minipage}
	\begin{minipage}[t]{0.33\linewidth}
		\centering
    	\includegraphics[width=1\textwidth]{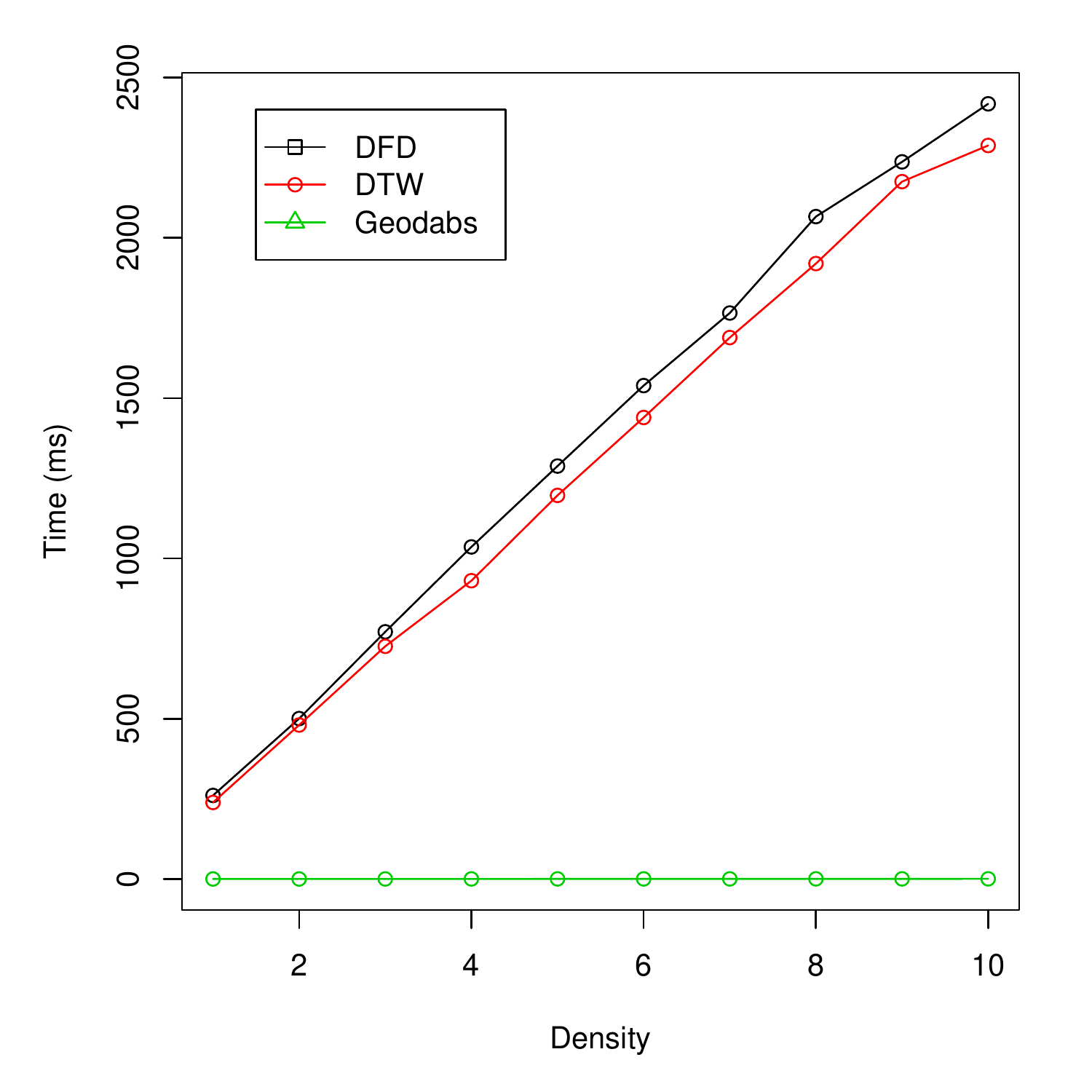}
    	\caption{Increasing the number \\ of trajectory candidates}
    	\label{fig:distance-limit}
	\end{minipage}
	\begin{minipage}[t]{0.33\linewidth}
		\centering
    	\includegraphics[width=1\textwidth]{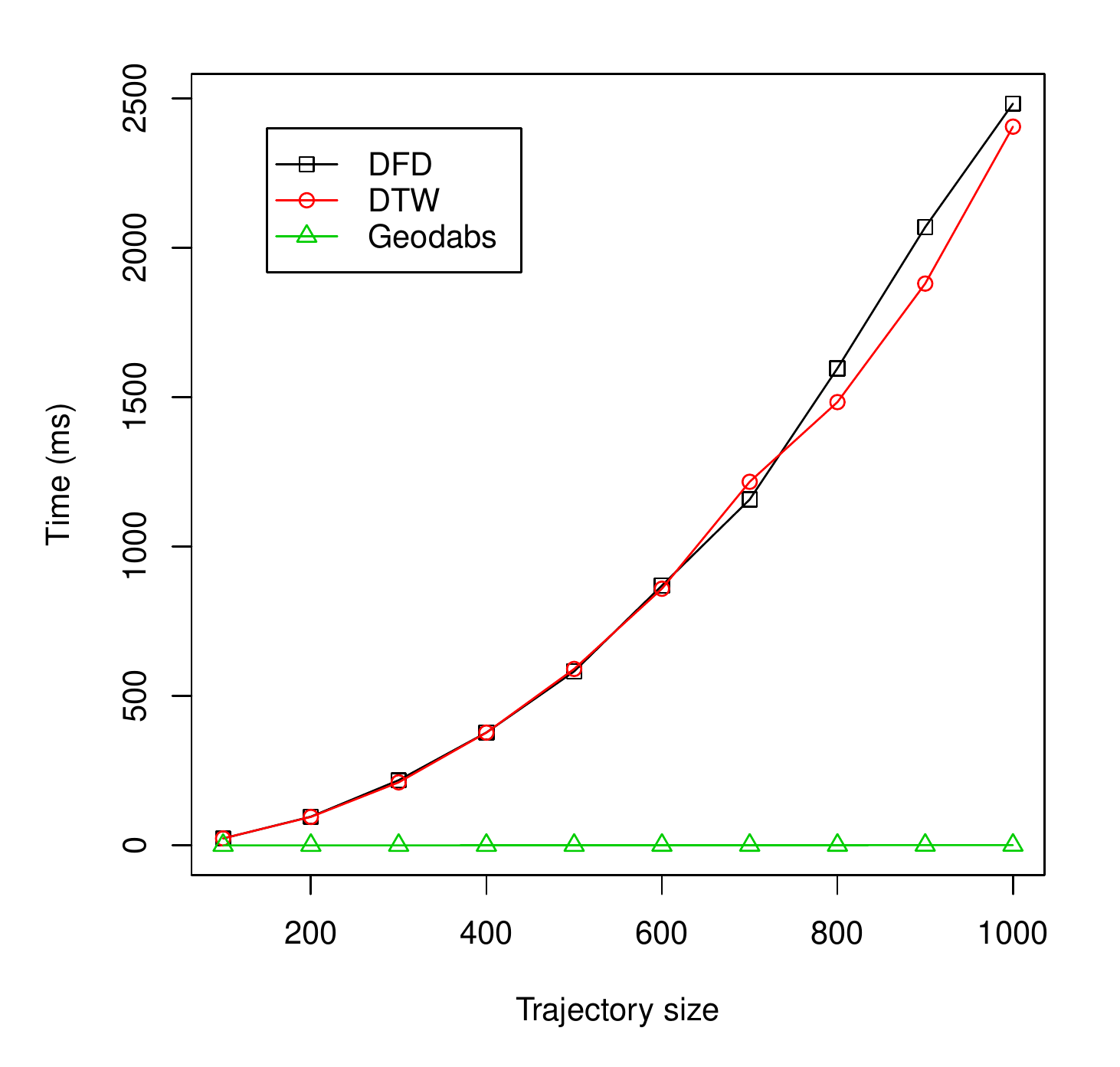}
    	\caption{Increasing the length \\ of the trajectory candidates}
    	\label{fig:distance-density}
	\end{minipage}
\end{figure*}

In this section, we highlight the cost of computing distances and discovering motifs with DFD, DTW and Jaccard in dense trajectory datasets.
To this aim, we characterize a large synthetic datasets and the configuration parameters we used to perform our evaluation.
Our experiments confirm the pragmatic observation made in section \label{sec:motivations} and enable us to focus on Jaccard based methods.
We then compare geodabs with geohashes, both in term of efficiency and effectiveness on a large and dense trajectory dataset.
Finally, we evaluate how a geodab index can be sharded across a set of nodes.

\subsection{Evaluation Setup}

\subsubsection{Datasets}

To perform our evaluation, we extensively rely on a synthetic dataset and on the road network extracted from the OpenStreetMap dataset~\cite{haklay2008openstreetmap}.
In the context of trajectory indexing, we noticed a lack of large and dense trajectory datasets.
For example, the Nokia and Geolife datasets~\cite{laurila2012mobile,zheng2010geolife} are too small and too sparse to validate our contributions.
In addition, these datasets lack the trajectory queries and the associated ground truth that is required to qualitatively assess our solution.
These kinds of queries and ground truths, also lack in the popular BerlinMod synthetic dataset~\cite{duntgen2009berlinmod}.
Therefore, we used the trajectory generator introduced in~\cite{chapuis2017scaling} to create a synthetic dataset, a set of trajectory queries and the associated ground truth.

Our dataset is based on 5'000 unique routes constrained on a road network and generated with the GraphHopper library~\cite{karich2014graphhopper}.
The routes are all located in a dense area of $300$ square kilometres located around the center of London.
We use these routes to generate $10$ similar trajectories in one direction and $10$ similar trajectories in the opposite direction.
These trajectories are sampled uniformaly at a rate of one point every second.
The speed of the moving entities is based on the route duration computed by the GraphHopper library.
In addition, we add $20$ meters of random Gaussian noise to every sampled points to make the dataset more realistic.

Figure \ref{fig:dataset} visually depicts the $5'000$ routes used to generate the trajectory dataset.
The $100'000$ trajectories derived from these routes form a very dense dataset of $6.3Gb$ that mimics the traces GPS trackers would record.
As of today, we found no publicly available counter parts to this synthetic dataset.
However, we believe that additional work on the generation of a dense trajectory dataset would greatly benefit the research community.
Figure \ref{fig:dataset-excerpt} illustrates a set of $10$ similar trajectories generated on the basis of a route constrained to the road network.

\subsubsection{Configuration Parameters}
\label{sec:configuration}

In order to evaluate the effectiveness of our solution, we need to find appropriate configuration parameters.
First, we empirically evaluated several configurations and observed the best results with a normalization based on geohashes of $36$ bits, a lower bound of $k=6$ and an upper-bound of $t=12$.
As we get closer to the poles, the width of the geohashes tends to shrink. 
In London, a geohash of $36$ bits has a width of $95$ meters and a height of $76$ meters.
As a trajectory normalized with geohashes rarely follows a diagonal path, we can assume that the average length of a move between two geohashes is approximately $85$ meters.
Therefore, the lower-bound $k$ translates to a segment threshold of approximately $510$ meters.
Segments shorter than this threshold are considered as noise.
The upper-bound $t$ translates to a segment threshold of approximately $1020$ meters.
Segments greater than this threshold are guaranteed to be detected. 
To validate our parameters, we tested several levels of normalization, performed  queries on a sample of our dataset and plotted the corresponding precision and recall curves.
As highlighted in Figure \ref{fig:config}, the normalization based on geohashes of $36$ bits clearly outperform its upstream and downstream counterparts.
Automating the discovery of the appropriate parameters is a difficult task, because the number of possible combinations is very large and each configuration requires building and querying an index.
A hill-climbing strategy could probably be used to address this problem, and this might be part of our future work.

\subsection{The Cost of Computing Distances}
\label{sec:computing-distances}

In this section, we characterize the cost of computing distances between trajectories and discuss their limits when searching trajectories in dense datasets.
We begin by reminding some fundamental distance measures.
We then compare them with the Jaccard distance used in the context of our paper.

\subsubsection{Ground Distance}

Equation~\ref{eq:dist} describes the haversine ground distance formula, where $R$ corresponds to the earth's radius in meters. 
Given a pair of points $p_l = (\varphi_l, \lambda_l)$ and $p_k = (\varphi_k, \lambda_k)$, the resulting value in $d(p_l, p_k)$ corresponds to the ground distance in meters.

\small
\begin{equation}
	\label{eq:dist}
	2 R \arcsin{\sqrt{\sin^{2}\left({\frac{\varphi_{l}-\varphi_{k}}{2}}\right)+\cos(\varphi_{k})\cos(\varphi_{l})\sin^{2}\left({\frac{\lambda_{l}-\lambda_{k}}{2}}\right)}}
\end{equation}
\normalsize

\subsubsection{Dynamic Time Warping}

Equation \ref{eq:dtw} presents the recursive function used to compute the dynamic time-warping distance (DTW)~\cite{yi1998efficient}.
Given a trajectory $P = \langle p_1, ..., p_m \rangle$ and a trajectory $Q = \langle q_1, ..., q_n \rangle$, the resulting value in $dtw(|P|,|Q|)$ corresponds to  the DTW distance between the trajectories.

\small
\begin{equation}
	\label{eq:dtw}
	dtw(i, j) = 
	\begin{cases}
		\infty & \text{if $i = 0$ or $j = 0$} \\
		0 & \text{if $i = j = 0$} \\
		d(P_i, Q_j) + min
		\begin{cases}
			dtw(i-1,j) \\
			dtw(i,j-1) \\
			dtw(i-1,j-1)
		\end{cases} & \text{otherwise}
	\end{cases}
\end{equation}
\normalsize

\subsubsection{Discrete Fréchet Distance}

Similarly, Equation \ref{eq:dfd} describes the recursive function used to compute the discrete Fréchet distance (DFD)~\cite{eiter1994computing}ite.
The resulting value in $dfd(|P|, |Q|)$ also corresponds to the DFD distance between the trajectories.

\small
\begin{equation}
	\label{eq:dfd}
	dfd(i, j) = 
	\begin{cases}
		d(P_i, Q_j) & \text{if $i = j = 1$} \\
		max 
		\begin{cases}
			d(P_i, Q_j)  \\
			min 
			\begin{cases}
				dfd(i-1,j) \\
				dfd(i,j-1) \\
				dfd(i-1,j-1)
			\end{cases}
		\end{cases} & \text{otherwise}
	\end{cases}
\end{equation}
\normalsize

\subsubsection{Performance Evaluation}

We compute the distance between a single query trajectory of length $t$ and a set of trajectory candidates of size $c$, where each candidate has a length of $t$.
In such a scenario, the computational cost associated with the computation of DTW and DFD is characterised by a complexity of $O(c*t^2)$.
In Figure~\ref{fig:distance-limit}, the size of the set of trajectory candidates $c$ remains constant, and the length of the query and candidate trajectories increases.
As highlighted here, as the length of the trajectories increases, so does the computational time in a polynomial manner.
Therefore, when a dataset is primarily made of long trajectories recorded at a high sampling rate, computing DTW or DFD is impractical.
In Figure~\ref{fig:distance-density}, the size of the set of trajectory candidates $c$ densifies and the length of the trajectories $t$ remains constant.
As the size of the of the candidate set increases, so does the computational time in a linear manner.
In both cases, we notice that computing the scores associated with $10$ trajectories of $1'000$ points takes more than $2500$ milliseconds.
In the context of a very dense trajectory dataset, a query can return many more relevant trajectory candidates, for which a distance measure still has to be computed.

Therefore, it is obvious that relying on these distance measures might be qualitatively sound but clearly unsustainable at scale.
In contrast, as highlighted in Figures~\ref{fig:distance-limit}~and~\ref{fig:distance-density}, computing the Jaccard distance for ordered sets of geodabs extracted from the trajectories is very inexpensive.
This clearly confirms the correctness of the pragmatic observation made in section~\ref{sec:motivations}.

\begin{figure*}[t!]
	\captionsetup{justification=centering}
	\begin{minipage}[t]{0.33\linewidth}
		\centering
    	\includegraphics[width=1\textwidth]{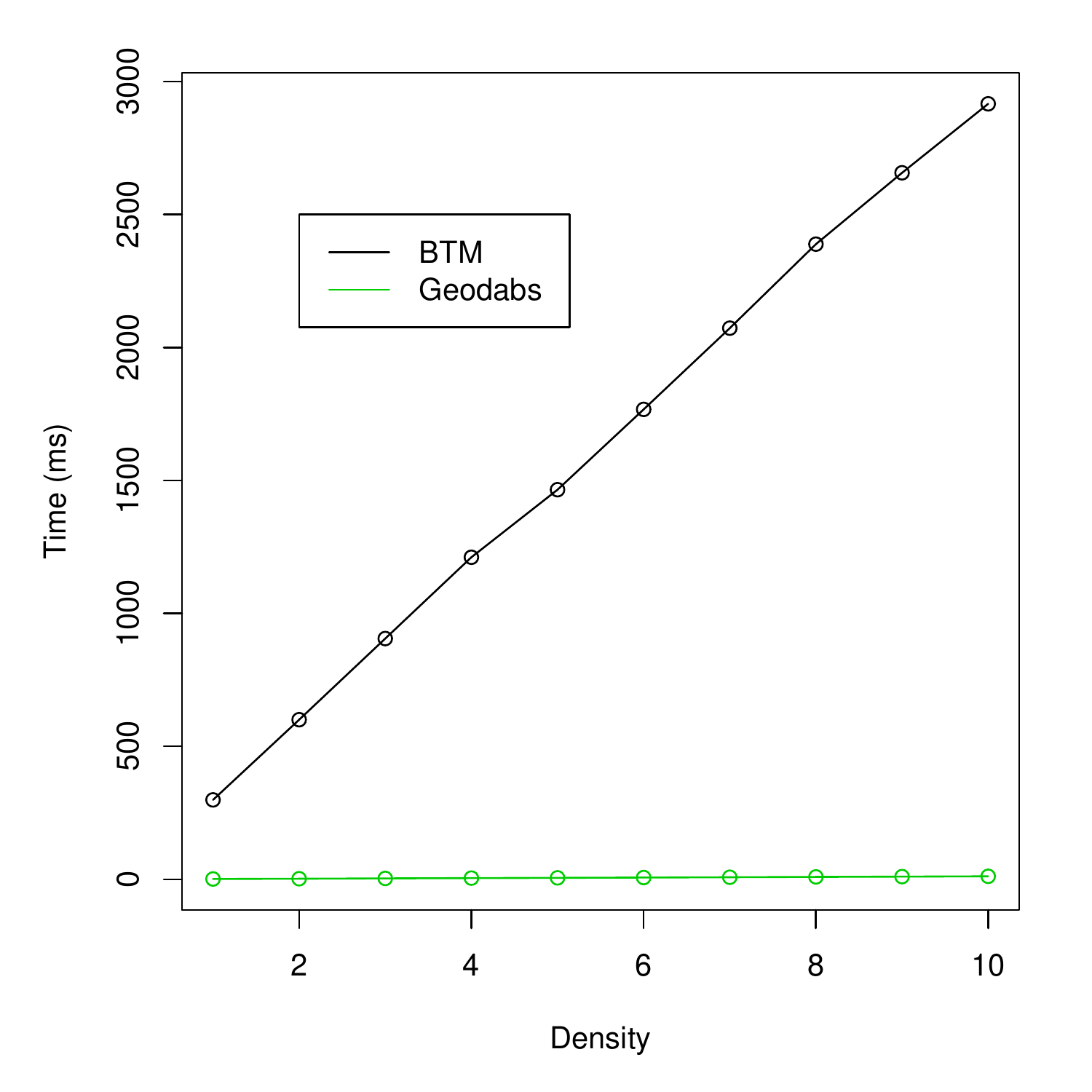}
    	\caption{Motif discovery with increasing trajectory candidates}
    	\label{fig:discovery}
	\end{minipage}
	\begin{minipage}[t]{0.33\linewidth}
		\centering
    	\includegraphics[width=1\textwidth]{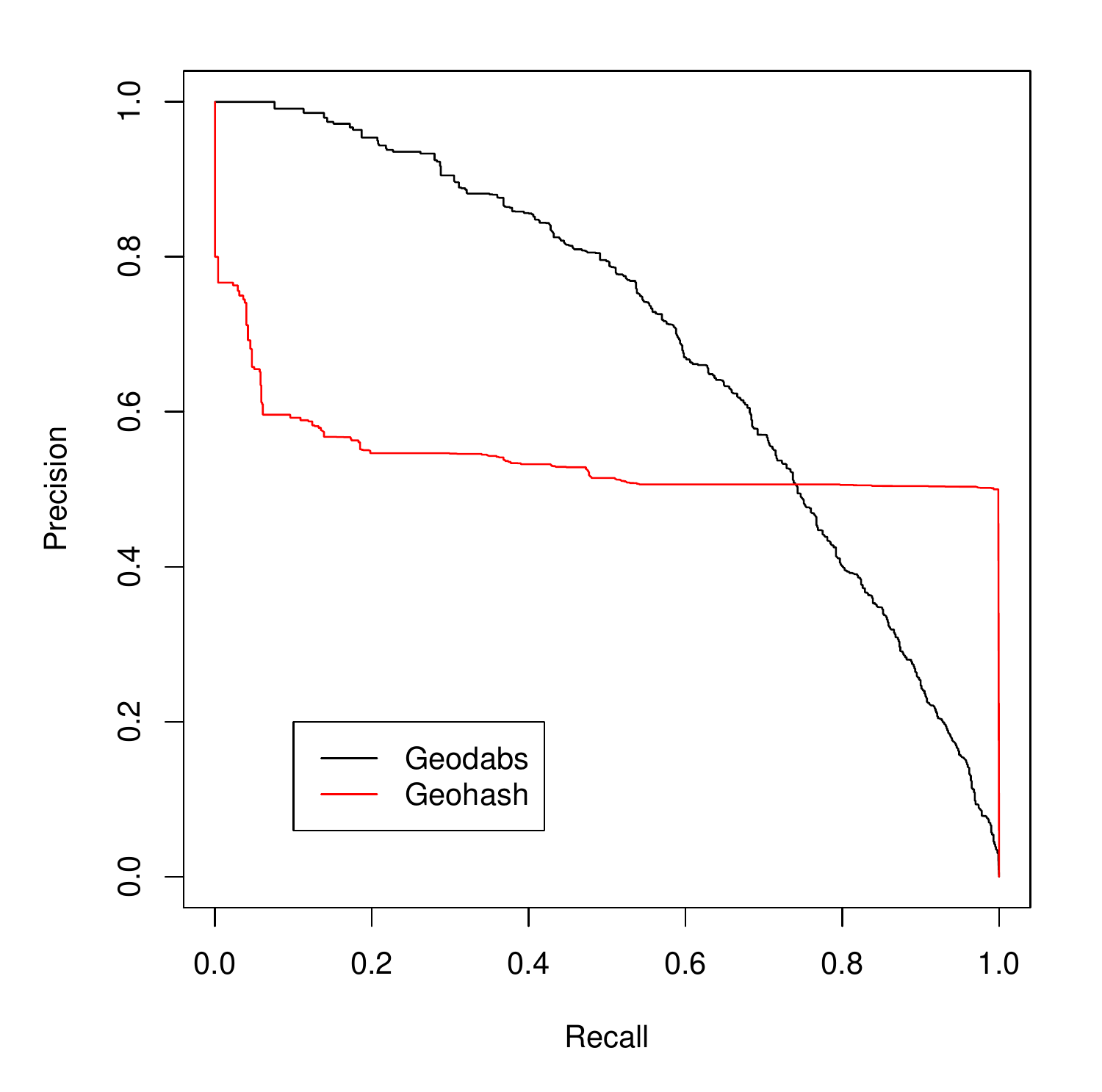}
    	\caption{PR curve}
    	\label{fig:pr-curve}
	\end{minipage}
	\begin{minipage}[t]{0.33\linewidth}
		\centering
    	\includegraphics[width=1\textwidth]{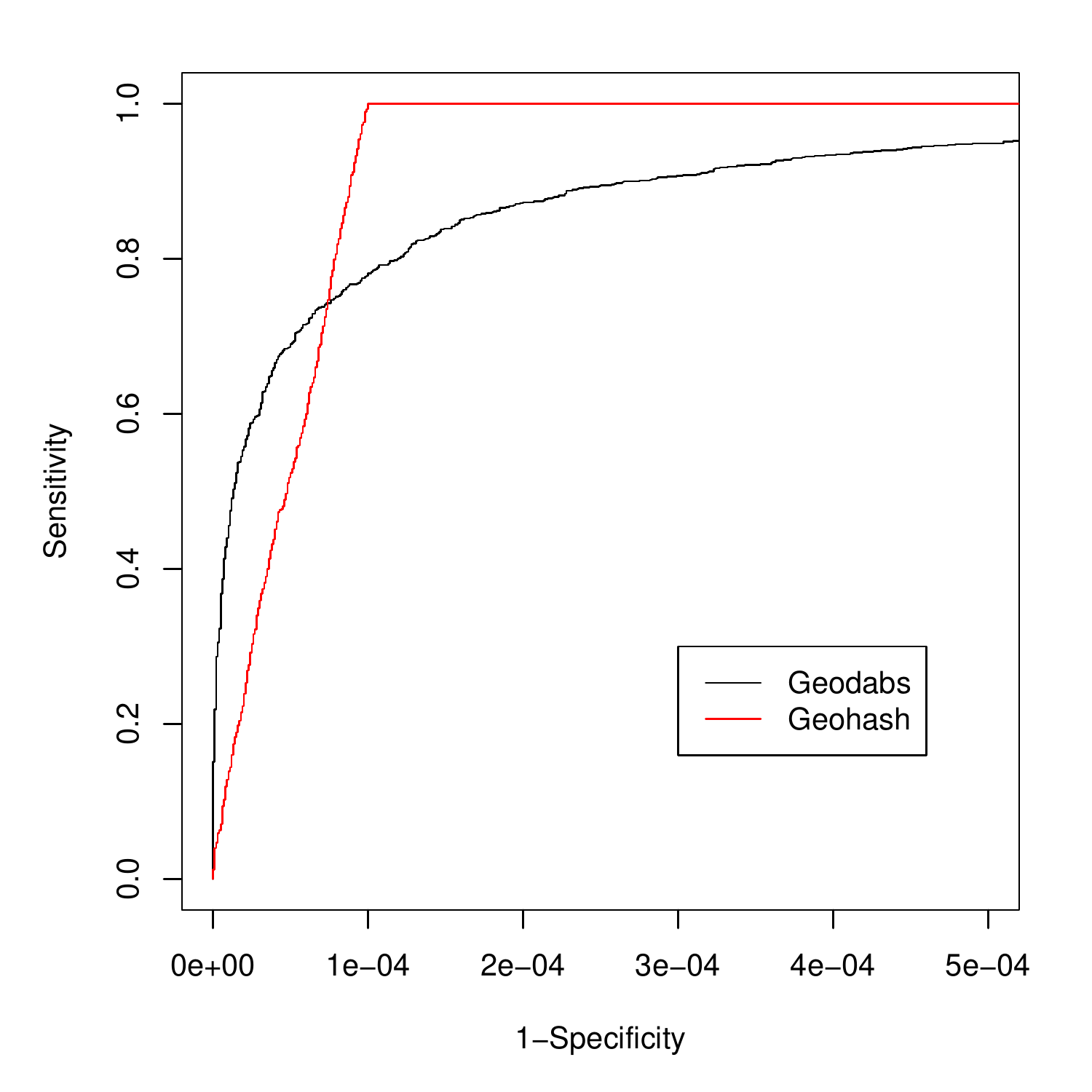}
    	\caption{ROC curve}
    	\label{fig:roc-curve}
	\end{minipage}
\end{figure*}

\subsection{The Cost of Discovering Motifs}

In order to find motifs in pairs of trajectories with geodabs, we have to make some assumptions regarding the normalization step.
First, we use our dataset to estimate the average number of fingerprints extracted per meters $a$ from normalized trajectories.
As a result, when looking for motifs of length $l$, we can translate this length to a number of fingerprints $f = l * a$.
Therefore, given two ordered sets of geodabs $F_i$ and $F_j$ 
obtained by fingerprinting the trajectories $S_i$ and $S_j$,
the problem now consists in returning a pair of motifs $(\bar{F}_i,\bar{F}_j)$ such that 
$\mathit{length}(\bar{F}_i) = \mathit{length}(\bar{F}_j) = f~\wedge \not\exists (\bar{F'}_i,\bar{F'}_j)$
for which 
$d_J(\bar{F'}_i,\bar{F'}_j)~<~d_J(\bar{F}_i,\bar{F}_j)$.
As the ordered sets $F_i$ and $F_j$ are usually relatively small, a brute force implementation of this method gives good results.
Because of the normalization and the fingerprinting, the motifs discovered with this approach are not strictly equivalent in terms of length and are subject to threshold effects.
However, the results we observed in practice are good approximations of the best result.

In Figure \ref{fig:discovery}, we compare our method with an optimized algorithm, called bounding-based trajectory motif (BTM), which gives exact solutions to the motif-discovery problem by computing DFD for every motif pair in the trajectories~\cite{tang2017efficient}.
As illustrated here, as the number of trajectory candidates densifies, so does the computational time.
Again, our method based on geodabs appears to be a necessary tradeoff.

\subsection{The Cost of Indiscrimination}

An inability of the index to discriminate between true and false positive translates to a greater set of trajectories for which the distance has to be computed.
In this section, we characterize the effectiveness and the probabilistic nature of the geohash and geodabs indexes.
We then show how an inability to discriminate directly affects performances.

\begin{figure*}[t!]
	\captionsetup{justification=centering}
	\begin{minipage}[t]{0.33\linewidth}
		\centering
    	\includegraphics[width=1\textwidth]{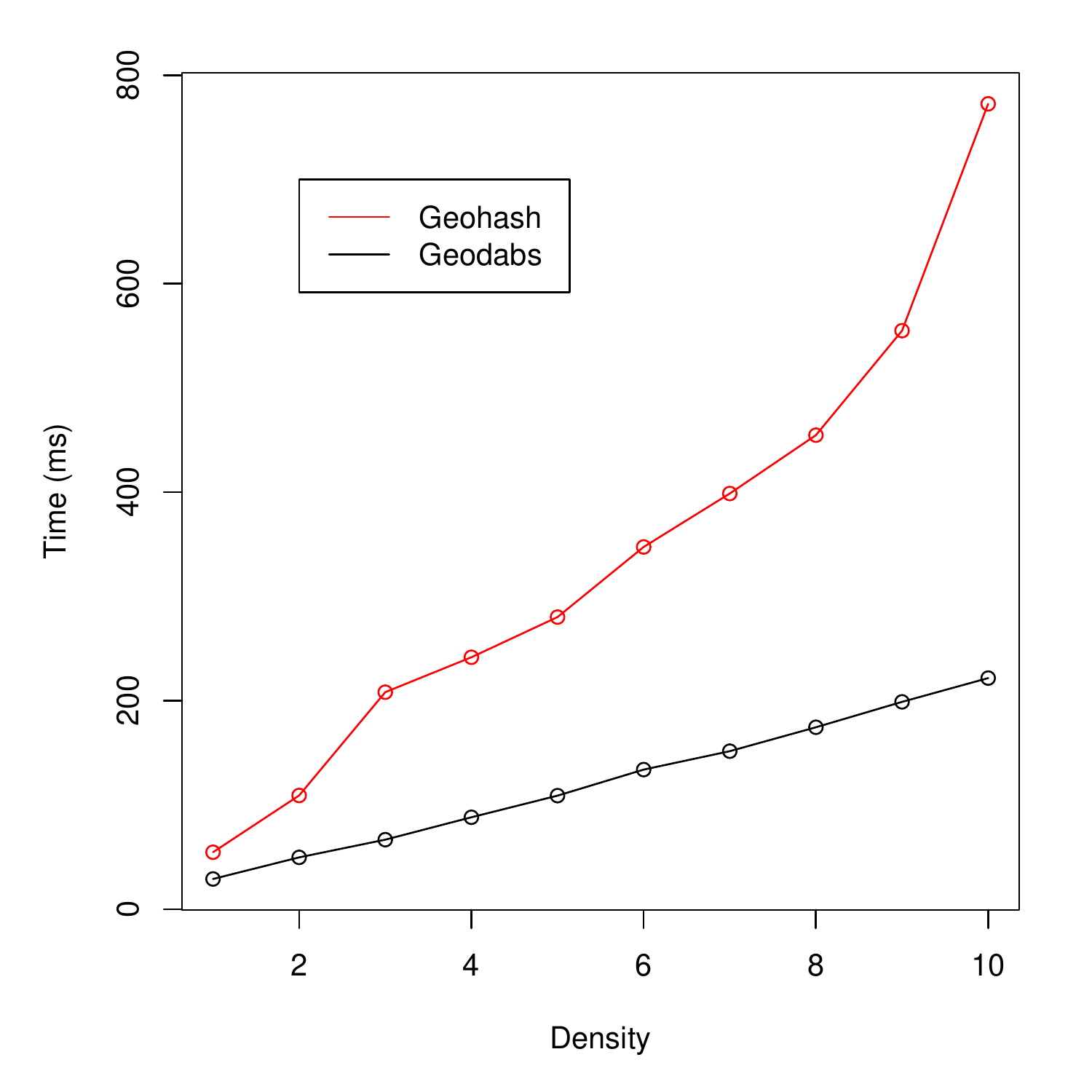}
    	\caption{Executing 100 queries on a large dataset of increasing density}
    	\label{fig:performances}
	\end{minipage}
	\begin{minipage}[t]{0.33\linewidth}
		\centering
    	\includegraphics[width=1\textwidth]{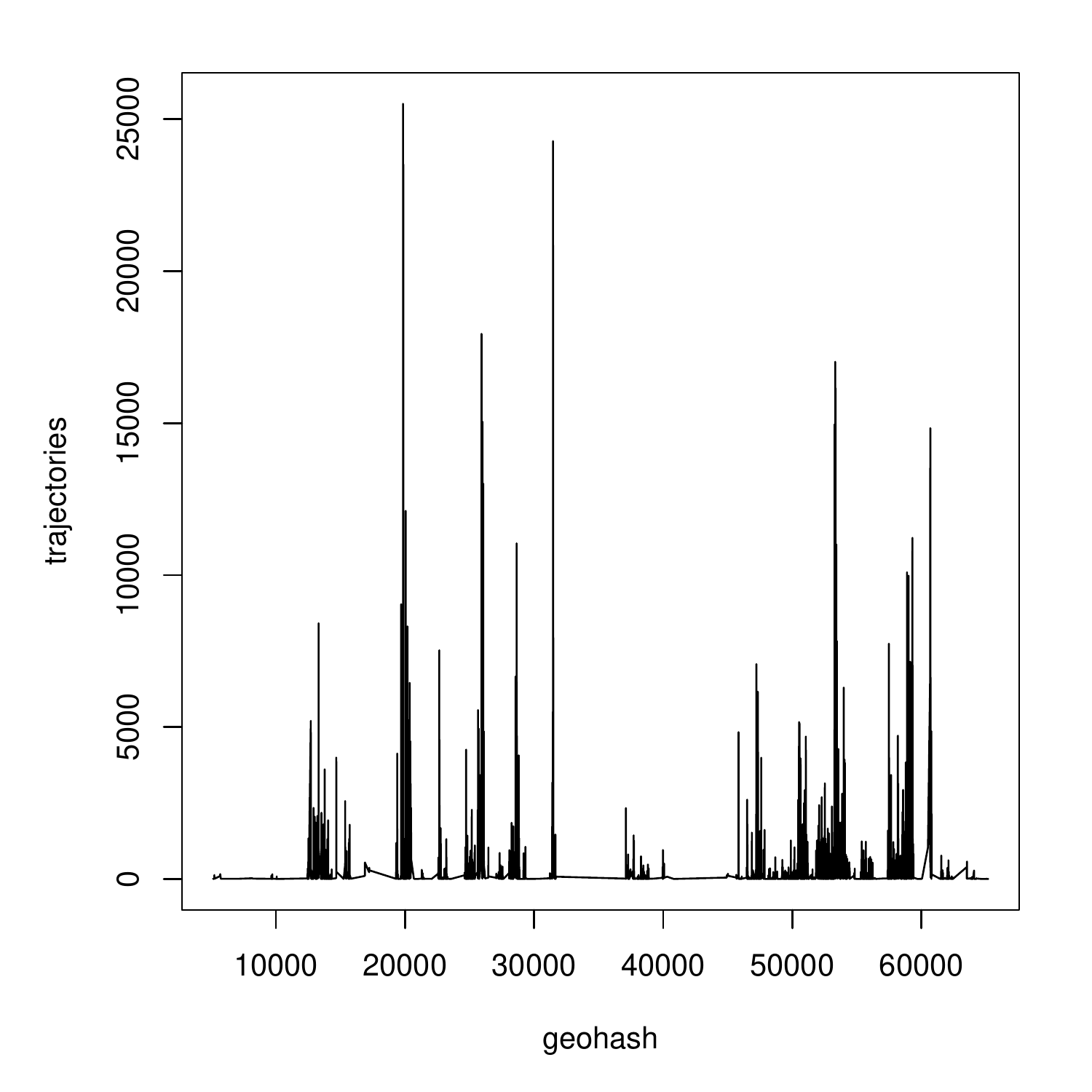}
    	\caption{Distribution of the trajectories \\ in geohash areas}
    	\label{fig:geodabs-distribution}
	\end{minipage}
	\begin{minipage}[t]{0.33\linewidth}
		\centering
    	\includegraphics[width=1\textwidth]{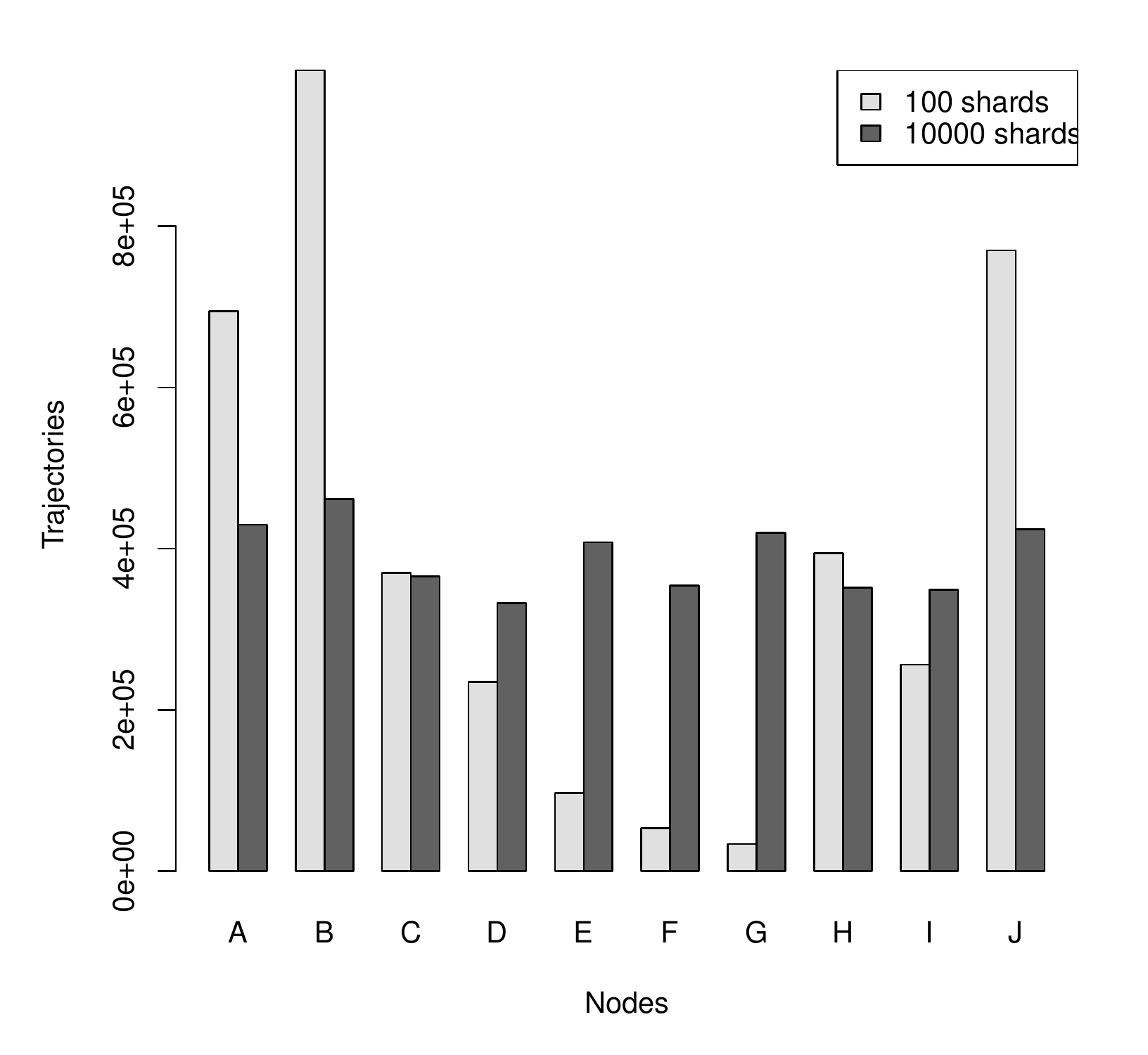}
    	\caption{Distribution of the trajectories \\ in a 10 nodes cluster}
    	\label{fig:nodes-distribution}
	\end{minipage}
\end{figure*}

\subsubsection{Index Effectiveness}

Figure~\ref{fig:pr-curve} depicts the PR curves obtained by querying the geodabs and geohash indexes~\cite{Manning:2008:IIR:1394399}.
When looking at the geohash curve, we first notice that precision drops rapidly as recall increases.
This is due to the inability of the geohash index to discriminate among similar trajectories that go in opposite directions.
Because each trajectory of our synthetic dataset is associated with a return path, the geohash curve tends to stabilise at a precision of~$0.5$, as recall increases.
The curve associated with the geodab index clearly shows that our method addresses this discrimination issue.
Furthermore, we also notice that the first results returned by the geodab index are characterized by very high precision.
In the context of a very dense dataset, this property is desirable because we can focus on the subset of the most relevant results.

Figure~\ref{fig:roc-curve} depicts the receiver-operating-characteristics (ROC) curve obtained by querying the geodab and geohash indexes~\cite{fawcett2004roc}.
In ranked retrieval, \textit{sensitivity} usually corresponds to \textit{recall} and \textit{specificity} is given by $tn/(fp+tn)$.
Thus, in contrast with the PR curve, the ROC curve enables us to qualitatively assess the full retrieval spectrum~\cite{Manning:2008:IIR:1394399}.
As we look at the full retrieval spectrum, the quality our results is exacerbated by the size of the dataset.
Therefore, it is important to notice that the plot focuses on a very narrow interval of the \textit{specificity}.
In fact, qualitatively speaking, both indexes are characterized by a very high sensitivity and a marginally low number of relevant results are lost.
This is confirmed by computing the area under the ROC curve (AUC) that is of $0.999889$ for geodabs and $0.9999521$ for geohashes.
Here, the minor difference in terms of AUC comes from the fact that a marginal number of relevant results can be missed with geodabs.
the fact that the curve associated with geodabs climbs more steeply, however, confirms that the first results returned by our method are more relevant.

\subsubsection{Index Efficiency}

Figure \ref{fig:performances} compares the average time needed to process $100$ queries on inverted indexes built with a sample of up to $10'000$ trajectories.
Here, in contrast with the results highlighted in Section~\ref{sec:computing-distances}, the number of trajectory candidates is not controlled.
Therefore, the difference between geohash and geodabs mainly highlights the inability of geohash to discriminate among trajectories. 
By combining several cells into one hash, a geodab not only discriminates on the direction of the trajectory, but also by all its constituents.
Therefore, the number of candidates for which the Jaccard distance has to be computed is significantly reduced.
As a result, processing queries is significantly faster but as shown earlier, the quality is not compromised.

\subsection{The Distribution of the Index}

We test the distributed nature of our index with a global road network extracted from the full dump of OpenStreetMap.
Here, we make the assumption that the distribution of trajectories recorded across the world should mostly fit on a road network and be characterized by a similar distribution.
The geodabs produced by our algorithm are characterized by a geohash prefix of $16$ bits that can easily be extracted with a bitwise operation.
Geohashes of depth $16$ subdivide space into $2^{16}$ cells characterized by a width of approximately 156 kilometres at the equator.
In Figure~\ref{fig:geodabs-distribution}, we plot the number of trajectories per geohash and we notice some very dense areas.
For example, the highest peak located on the left of the diagram corresponds to the geohashes located around Mexico city.
In contrast, the voids we have between the peaks correspond to areas of low activity, such as oceans.
In Figure \ref{fig:nodes-distribution}, we assume that the index is distributed across $10$ nodes.
We notice that a small number of shards ($100$) is not sufficient to distribute the data in a balanced fashion.
However, with a greater number of shards ($10'000$), the data are well balanced on the nodes.
Therefore, there is a tradeoff to make between preserving locality (to reduce the number of shards contacted when performing queries) and breaking locality (to spread the data evenly in the the cluster).

\section{Conclusion}

In this paper, we have shown that fingerprinting, and more specifically winnowing, can be used for indexing trajectories.
We have introduced geodabs, a construction that combines hashing and geohashing to discriminate on the spatial and on the temporal dimensions.
In addition, we have shown that geodabs can be used to scale and distribute an index across several nodes in a cluster.
We have demonstrated how trajectory normalization can be used to improve the quality of an index.
Finally, we discussed several pragmatic experiments that demonstrated the effectiveness and efficiently of trajectory fingerprinting with geodabs.

\bibliographystyle{abbrv}
\bibliography{paper.bib}

\begin{thebibliography}{10}

\bibitem{bentley1975multidimensional}
J.~L. Bentley.
\newblock Multidimensional binary search trees used for associative searching.
\newblock {\em Communications of the ACM}, 18(9):509--517, 1975.

\bibitem{brin1995copy}
S.~Brin, J.~Davis, and H.~Garcia-Molina.
\newblock Copy detection mechanisms for digital documents.
\newblock In {\em ACM SIGMOD Record}, volume~24, pages 398--409. Acm, 1995.

\bibitem{broder1997resemblance}
A.~Z. Broder.
\newblock On the resemblance and containment of documents.
\newblock In {\em Compression and Complexity of Sequences 1997. Proceedings},
  pages 21--29. IEEE, 1997.

\bibitem{chakka2003indexing}
V.~P. Chakka, A.~C. Everspaugh, and J.~M. Patel.
\newblock Indexing large trajectory data sets with seti.
\newblock {\em Ann Arbor}, 1001(48109-2122):12, 2003.

\bibitem{chapuis2017dss}
B.~Chapuis, P.~Andritsos, and B.~Garbinato.
\newblock An efficient type-agnostic approach for finding sub-sequences in
  data.
\newblock In {\em Data Science and Systems (DSS), 2017 3rd International
  Conference on}. IEEE, 2017.

\bibitem{chapuis2017scaling}
B.~Chapuis and B.~Garbinato.
\newblock Scaling and load testing location-based publish and subscribe.
\newblock In {\em Distributed Computing Systems (ICDCS), 2017 IEEE 37th
  International Conference on}, pages 2543--2546. IEEE, 2017.

\bibitem{dewan2017som}
P.~Dewan, R.~Ganti, and M.~Srivatsa.
\newblock Som-tc: Self-organizing map for hierarchical trajectory clustering.
\newblock In {\em Distributed Computing Systems (ICDCS), 2017 IEEE 37th
  International Conference on}, pages 1042--1052. IEEE, 2017.

\bibitem{duntgen2009berlinmod}
C.~D{\"u}ntgen, T.~Behr, and R.~H. G{\"u}ting.
\newblock Berlinmod: a benchmark for moving object databases.
\newblock {\em The VLDB Journal—The International Journal on Very Large Data
  Bases}, 18(6):1335--1368, 2009.

\bibitem{eiter1994computing}
T.~Eiter and H.~Mannila.
\newblock Computing discrete fr{\'e}chet distance.
\newblock Technical report, Tech. Report CD-TR 94/64, Information Systems
  Department, Technical University of Vienna, 1994.

\bibitem{fawcett2004roc}
T.~Fawcett.
\newblock Roc graphs: Notes and practical considerations for researchers.
\newblock {\em Machine learning}, 31(1):1--38, 2004.

\bibitem{finkel1974quad}
R.~A. Finkel and J.~L. Bentley.
\newblock Quad trees a data structure for retrieval on composite keys.
\newblock {\em Acta informatica}, 4(1):1--9, 1974.

\bibitem{goh2012online}
C.~Y. Goh, J.~Dauwels, N.~Mitrovic, M.~T. Asif, A.~Oran, and P.~Jaillet.
\newblock Online map-matching based on hidden markov model for real-time
  traffic sensing applications.
\newblock In {\em Intelligent Transportation Systems (ITSC), 2012 15th
  International IEEE Conference on}, pages 776--781. IEEE, 2012.

\bibitem{guttman1984r}
A.~Guttman.
\newblock {\em R-trees: A dynamic index structure for spatial searching},
  volume~14.
\newblock ACM, 1984.

\bibitem{haklay2008openstreetmap}
M.~Haklay and P.~Weber.
\newblock Openstreetmap: User-generated street maps.
\newblock {\em IEEE Pervasive Computing}, 7(4):12--18, 2008.

\bibitem{heintze1996scalable}
N.~Heintze et~al.
\newblock Scalable document fingerprinting.
\newblock In {\em 1996 USENIX workshop on electronic commerce}, volume~3, 1996.

\bibitem{karich2014graphhopper}
P.~Karich and S.~Schr{\"o}der.
\newblock Graphhopper.
\newblock {\em http://www.graphhopper.com, last accessed}, 4(2):15, 2014.

\bibitem{kosub2016note}
S.~Kosub.
\newblock A note on the triangle inequality for the jaccard distance.
\newblock {\em arXiv preprint arXiv:1612.02696}, 2016.

\bibitem{laurila2012mobile}
J.~K. Laurila, D.~Gatica-Perez, I.~Aad, O.~Bornet, T.-M.-T. Do, O.~Dousse,
  J.~Eberle, M.~Miettinen, et~al.
\newblock The mobile data challenge: Big data for mobile computing research.
\newblock In {\em Pervasive Computing}, number EPFL-CONF-192489, 2012.

\bibitem{lemire2017roaring}
D.~Lemire, O.~Kaser, N.~Kurz, L.~Deri, C.~O'Hara, F.~Saint-Jacques, and
  G.~Ssi-Yan-Kai.
\newblock Roaring bitmaps: Implementation of an optimized software library.
\newblock {\em arXiv preprint arXiv:1709.07821}, 2017.

\bibitem{manber1994finding}
U.~Manber et~al.
\newblock Finding similar files in a large file system.
\newblock In {\em Usenix Winter}, volume~94, pages 1--10, 1994.

\bibitem{Manning:2008:IIR:1394399}
C.~D. Manning, P.~Raghavan, and H.~Sch\"{u}tze.
\newblock {\em Introduction to Information Retrieval}.
\newblock Cambridge University Press, New York, NY, USA, 2008.

\bibitem{newson2009hidden}
P.~Newson and J.~Krumm.
\newblock Hidden markov map matching through noise and sparseness.
\newblock In {\em Proceedings of the 17th ACM SIGSPATIAL international
  conference on advances in geographic information systems}, pages 336--343.
  ACM, 2009.

\bibitem{niemeyer2008geohash}
G.~Niemeyer.
\newblock Geohash, 2008.

\bibitem{pfoser2000novel}
D.~Pfoser, C.~S. Jensen, Y.~Theodoridis, et~al.
\newblock Novel approaches to the indexing of moving object trajectories.
\newblock In {\em VLDB}, pages 395--406, 2000.

\bibitem{schleimer2003winnowing}
S.~Schleimer, D.~S. Wilkerson, and A.~Aiken.
\newblock Winnowing: local algorithms for document fingerprinting.
\newblock In {\em Proceedings of the 2003 ACM SIGMOD international conference
  on Management of data}, pages 76--85. ACM, 2003.

\bibitem{srivatsa2017limits}
M.~Srivatsa, R.~Ganti, and P.~Mohapatra.
\newblock On the limits of subsampling of location traces.
\newblock In {\em Distributed Computing Systems (ICDCS), 2017 IEEE 37th
  International Conference on}, pages 1032--1041. IEEE, 2017.

\bibitem{tang2017efficient}
B.~Tang, M.~L. Yiu, K.~Mouratidis, and K.~Wang.
\newblock Efficient motif discovery in spatial trajectories using discrete
  fr{\'e}chet distance.
\newblock EDBT, 2017.

\bibitem{yi1998efficient}
B.-K. Yi, H.~Jagadish, and C.~Faloutsos.
\newblock Efficient retrieval of similar time sequences under time warping.
\newblock In {\em Data Engineering, 1998. Proceedings., 14th International
  Conference on}, pages 201--208. IEEE, 1998.

\bibitem{zheng2010geolife}
Y.~Zheng, X.~Xie, and W.-Y. Ma.
\newblock Geolife: A collaborative social networking service among user,
  location and trajectory.
\newblock {\em IEEE Data Eng. Bull.}, 33(2):32--39, 2010.

\end{thebibliography}

\end{document}